\documentstyle[preprint,prb,aps,epsf]{revtex}
\def\ospace{\o\hbox{ }}
\def\pt{\partial}

\begin{document}
\draft

%ABSTRACT
\title{Instabilities in the Flux Line Lattice of Anisotropic
 Superconductors}

\author {A.M. Thompson and M.A. Moore}
\address{Theory Group, Department of Physics, University of
 Manchester,}
\address{Manchester, M13 9PL, U.K.}
\maketitle

\begin{abstract}
The stability of the flux line lattice has been investigated within
 anisotropic London theory.
This is the first full-scale investigation of instabilities in the 
 `chain' state, 
 the equilibrium lattice that is similar to the Abrikosov lattice at 
 large fields but crosses over smoothly to a pinstripe structure at 
 low fields. 
By calculating the normal modes of the elasticity matrix, it has been
 found the lattice is stable at large fields, but that instabilities
 occur as the field is reduced.
The field at which these instabilities first arise,
 $b^*(\epsilon,\theta)$, depends on the anisotropy $\epsilon$ and the
 angle $\theta$ at which the lattice is tilted away from the $c$-axis.
These instabilities initially occur at wavevector ${\bf
 k}^*(\epsilon,\theta)$.
The dependence of ${\bf k}^*$ on $\epsilon$ and $\theta$ is 
 complicated, but the component of ${\bf k}^*$ along the average
 direction of the flux lines, $k_z$, is always finite.
For rigid straight flux lines, the cutoff necessary for London theory
 has been `derived' from Landau-Ginzburg theory, where the shape of
 the vortex core is known.
However, for investigating instability at finite $k_z$ it is necessary
 to know the dependence of the cutoff on $k_z$, and we have used a
 cutoff suggested by Sudb\ospace and Brandt. 
The instabilities  only occur for values of the anisotropy $\epsilon$ 
 appropriate to a material like BSCCO, and not for anisotropies more 
 appropriate to YBCO. 
The lower critical field $H_{c_1}(\phi)$ is calculated as a function 
 of the angle $\phi$ at which the applied field is tilted away from 
 the crystal axis. 
The presence of kinks in $H_{c_1}(\phi)$ is seen to be related to 
 instabilities in the equilibrium flux line structure. 

\end{abstract}
\narrowtext

%%%%%%%%%%%%%%%%%%%%%%%%%%%%%%%%%%%%%%%%%%%%%%%%%%%%%%%%%%%%%%%%%%%%%%
% INTRODUCTION
%%%%%%%%%%%%%%%%%%%%%%%%%%%%%%%%%%%%%%%%%%%%%%%%%%%%%%%%%%%%%%%%%%%%%%

\section{Introduction}

The existence of novel flux line structures in high-temperature 
 superconductors has led to intensive investigation of the mixed-state
 of these materials.
Unusual structures have been observed in Bitter pattern experiments on
 YBCO\cite{gammel,grigorieva1} and BSCCO
 \cite{bolle,grigorieva2} when the applied magnetic field was
 tilted away from the $c$-axis.
In YBCO, the `chain' state was
 observed\cite{gammel,grigorieva1} where the anisotropy of the
 material causes the usual repulsive flux line interaction to 
become attractive within the tilt plane, the plane containing the
 magnetic field and the $c$-axis. 
The chain state had been predicted within the framework of the
 London approximation\cite{grishin,buzdin}. 
In some of the experiments\cite{grigorieva1} the chains of flux lines
 were seen embedded in an approximately triangular flux line lattice,
 but it is believed the presence of the lattice was due to pinning
 of the flux lines. 
In BSCCO, similar structures of chains embedded in a lattice were 
 also seen \cite{bolle,grigorieva2}, but the dependence of
 the flux line spacings on the tilt angle and magnetic field were
 different from those in YBCO, implying these structures may be
 created by a different mechanism.
Possible explanations for the flux line structure seen in BSCCO have
 been proposed\cite{huse,sudbobrandthuse}.
It was suggested there existed inter-penetrating flux line lattices,
 one orientated approximately parallel to the $c$-axis while the other
 is orientated approximately parallel to the ab-plane.
Within the framework of the London approximation, it has been shown
 that provided the anisotropy is large enough there is more than one
 possible angle at which the flux lines initially enter the
 sample\cite{sudbobrandthuse}.

The mixed state for isotropic superconductors is a periodic triangular
 array of straight flux lines\cite{abrikosov}.
In uniaxially anisotropic superconductors it was
 predicted\cite{campbell,petzinger} the flux lines would form a
 distorted triangular lattice, where the spacings between the flux
 lines depend on the strength of the magnetic field ${\bf B}$ and the
 anisotropy mass ratio $M_z / M$.
A lattice similar to this distorted triangular lattice has been 
 observed in YBCO at large fields using small angle neutron 
 scattering\cite{yethiraj}.

The stability of this distorted lattice against elastic deformations
 has been studied extensively within London theory.
Sudb\ospace and Brandt\cite{sudbobrandt} showed that at large 
 anisotropy$M_z/M \gg 1$ and small magnetic induction 
 $b=B/H_{c_2} \ll 1$ theenergy associated with a pure shearing mode 
 of the flux lattice canbecome negative.
The existence of a tilt-wave instability, ${\bf k}=(0,0,k_z)$, was
 demonstrated by Sardella and Moore\cite{sardellamoore} and confirmed
 by Nguyen and Sudb\ospace\cite{nguyensudbo}, who both employed the 
 same cutoff procedure. 
%new text
The distorted triangular lattice is the lattice one would expect 
 using the ideas of anisotropic scaling\cite{rescale}. 
The lattice's basis vectors are 
 proportional to $1/\sqrt{b}$ and depend in a simple manner on 
 $\theta$, $\kappa$, $\epsilon$. 
Minimizing the free energy, Daemen et al.\cite{koganlattice2} showed 
 that out of the set of centered rectangle lattice structures, 
 one of which is the distorted triangular lattice, the true 
 equilibrium lattice behaves quite differently. 
At large fields it is approximately the distorted triangular 
 (Abrikosov) lattice, but there is a smooth crossover to the `chain' 
 state at small fields. 
This state has one of the basis vectors independent of the field, 
 with the other being inversely proportional to $b$. 

In this paper we investigate the stability of this equilibrium 
 lattice at a general wavevector ${\bf k}=(k_x,k_y,k_z)$. 
The existence of a zone center instability has been 
 observed\cite{sardella2}, but this is the first full-scale 
 investigation of elastic instabilities of this equilibrium lattice. 
The cutoff used is that proposed by Sudb\ospace and 
 Brandt\cite{sudbobrandt}, which depends on $k_z$. 
The lattice is found to be stable at large fields. 
As the field is reduced, the field at which the instability first 
 occurs $b^*(\epsilon,\theta)$
 depends on the anisotropy $\epsilon$, and the angle $\theta$ at which
 the lattice is tilted from the c-axis.
The magnitude of these fields $b=B/H_{c_2} \sim O(10^{-4})$
 is approximately that used in the experiments where the unusual flux 
 line structures were seen.
It is generally believed these instabilities are somehow related to
 the different flux line structures seen in experiments, but we have
 as yet no good theoretical interpretation of the connection.

%Nonmono
The attractive vortex interaction also forces the flux lines to 
 initially enter a superconductor not as single flux lines but in 
 chains. 
For very large anisotropy, there exists the possibility that these 
 chains may first enter the sample at more than one direction. 
The precise details are complicated, see section \ref{sec:chain}. 
However, they indicate that for YBCO the chain state is the stable 
 low field structure. This is not the case for BSCCO, where it is 
 possible the low field structure will be the coexistence of 
 different chain  state orientations.

%cutoff
The initial instabilities observed always have {\it finite} $k_z$. 
For instabilities of non-zero $k_z$ the form of the cutoff used within
 the London theory is crucial. 
In the calculation of Nguyen and Sudb\ospace\cite{nguyensudbo}, which 
 used a cutoff that did not depend on $k_z$, the coexistence of 
 different flux line orientations and the zero field tilt wave 
 instability were two different effects, i.e. they initially occurred 
 at different anisotropies. 
In the limit $b \to 0$ it can be shown that the onset of both 
 effects are related to the line tension $P_l(\theta) = 
 \varepsilon_l(\theta) + \pt^2 \varepsilon_l /\pt \theta^2$ becoming 
 negative. If the cutoff that depends on $k_z$ is used, it is found 
 that the two effects do occur at the same anisotropy, implying this 
 cutoff may be more reliable. 

%%%%%%%%%%%%%%%%%%%%%%%%%%%%%%%%%%%%%%%%%%%%%%%%%%%%%%%%%%%%%%%%%%%%%%
% LONDON THEORY
%%%%%%%%%%%%%%%%%%%%%%%%%%%%%%%%%%%%%%%%%%%%%%%%%%%%%%%%%%%%%%%%%%%%%%
\section{London Theory}

A convenient way to describe the low-field magnetic properties of
 high-$T_c$ superconductors is London theory.
In the isotropic form, this theory just depends on the penetration 
 depth $\lambda$ and on $\kappa=\lambda/\xi$, where $\xi$ is the 
 coherence length. 
To allow for the anisotropy of the HTSC compounds, the square
 penetration depth $\lambda^2$ is replaced by the tensor
 $\lambda_{ij}^2 = \Lambda_{ij}$.
Here, we shall only investigate uniaxial anisotropy where
 $\Lambda_{XX}=\Lambda_{YY}=\lambda_{ab}^2\neq\Lambda_{ZZ}=
 \lambda_c^2$ are the only non-zero elements of $\Lambda_{ij}$.
The anisotropy of the material is governed by the parameter $\epsilon$
 which in the effective mass model is given by $\epsilon^2=
 {M_{XX}/M_Z}$, and $\lambda_{ab} / \lambda_{c}= \xi_{c} / \xi_{ab} 
 = \epsilon$.

The London free energy can be written as
\begin{equation}
F={1\over 8\pi} \int d^3r \left\{ {\bf H}^2 + \left( {\Phi_0\over
 2\pi}\nabla\varphi - {\bf A} \right) \cdot {\bf \Lambda}^{-1} 
 \cdot \left( {\Phi_0\over 2\pi}\nabla\varphi - {\bf A} \right) 
 \right\}
 \label{londonfree}
\end{equation}
where ${\bf A}$ is the vector potential of the magnetic field ${\bf
 H}=\nabla\times {\bf A}$, $\varphi$ is the phase of the order
 parameter, $\Phi_0$ is the flux quantum, and ${\bf \Lambda}^{-1}$ 
 is the inverse of the square penetration depth tensor 
 ${\bf \Lambda}$.
The magnetic induction ${\bf B}$ is the average magnetic field ${\bf
 B}=\langle {\bf H}\rangle = B\hat{\bf z}$. London theory is a good
 approximation at low induction, $B < 0.2 H_{c_2}$\cite{brandt}, where
 the cores do not overlap strongly.

In general, the ${\bf B}$ field is not aligned with the crystal axis,
 and we chose our coordinate system such that the ${\bf B}$ field lies
 in the $X-Z$ plane and is tilted away from the c-axis by an angle
 $\theta$.
It is then convenient to use the `vortex' coordinate system
 (xyz). This is obtained by rotating the crystal frame (XYZ) by an
 angle $\theta$ around the Y-axis, see Fig. 1.
In the vortex coordinate system, the square penetration depth is given
 by
$\Lambda_{\alpha\beta} = \Lambda_1\delta_{\alpha\beta} + \Lambda_2
 c_\alpha c_\beta $ where $\Lambda_1=\lambda_{ab}^2$,
 $\Lambda_2=\lambda_c^2 - \lambda_{ab}^2$, $(\alpha, \beta)=
 (x,y,z)$, and $c_\alpha$ is the $\alpha$ component of the unit 
 vector $\hat {\bf c}$ in the vortex frame.

The free energy (\ref{londonfree}) can be written in a
 simpler form. Minimizing (\ref{londonfree}) with respect to the
 vector potential ${\bf A}$, and then taking the curl of the equation,
 we obtain the London equation

\begin{equation}
{\bf H} + \nabla\times \left\{ {\bf \Lambda} \cdot \nabla\times 
 {\bf H} \right\} = \Phi_0 \sum_i \int \hbox{d}{\bf r}_i \delta_3
 \left( {\bf r}-{\bf r}_i\right) \label{loneqn}
\end{equation}
It is possible to derive this equation from the Ginzburg-Landau
 equations, assuming the order parameter has constant magnitude.
The right hand side of the London equation comes from
 $\nabla\times\nabla\varphi = \sum_i \int \hbox{d}{\bf r}_i
 \delta_3\left( {\bf r}-{\bf r}_i\right) $, where ${\bf r}_i$ is the
 position of the $i^{\hbox{th}}$ flux line. 

The London equation is linear, and has the solution
\begin{equation}
{\bf H}_\alpha\left( {\bf r} \right) = \Phi_0 \sum_i \int \hbox{d}{\bf
 r}_i V_{\alpha\beta}\left( {\bf r}-{\bf r}_i \right)
\end{equation}
Using this potential, the London free energy (\ref{londonfree}) may
 now be written as
\begin{equation}
F={\Phi_0^2 \over 8 \pi} \sum_{i,j} \int\int \hbox{d}{\bf r}_i^\alpha
 \hbox{d}{\bf r}_j^\beta V_{\alpha\beta}\left( {\bf r}_i - {\bf r}_j
 \right) \label{londonfree2}
\end{equation}
The convention of summation over repeated indices is assumed.
Written in this form, the free energy can be seen as consisting of two
 parts, the self-energy terms (~$~i~=~j~$~) and the interaction terms
 ($i\neq j$).
% cutoff....

The Fourier transform of the potential $V({\bf r} - {\bf r}_i)$ is
 \cite{barfordgunn}
\begin{equation}
V_{\alpha\beta}\left({\bf k}\right) = {1 \over 1 + \Lambda_1 k^2}
 \left[ \delta_{\alpha\beta} - {\Lambda_2 q_\alpha q_\beta \over 1 +
 \Lambda_1 k^2 + \Lambda_2 q^2 } \right] \label{potential}
\end{equation}
where ${\bf q}={\bf k} \times \hat{\bf c}$.
From Eq. (\ref{potential}), we see that the potential decays only
 as $1/k^2$ as $k\to\infty$, implying that ${\bf H}\left({\bf
 r}\right)$ is singular at ${\bf r}={\bf r}_i$.
The singularites in ${\bf H} \left({\bf r}\right)$ are due to the
 absence of the vortex cores from London theory.
A convenient way to circumvent the problems associated with the
 divergences is to introduce a cutoff into the London potential
 (\ref{potential}) via
\begin{equation}
V_{\alpha\beta}\left({\bf k}\right) = {S({\bf k}) \over 1 + \Lambda_1
 k^2} \left[ \delta_{\alpha\beta} - {\Lambda_2 q_\alpha q_\beta \over
 1 + \Lambda_1 k^2 + \Lambda_2 q^2 } \right] \label{potential2}
\end{equation}
This is equivalent to replacing the delta function in the London
 equation (\ref{loneqn}) with a short-ranged function $S\left({\bf
 r}-{\bf r}_i\right)$.
By making this short-ranged function a gaussian, of width $\sqrt{2
 \xi_{ab}}$ along $a$ and $b$ and width $\sqrt{2 \xi_c}$ along $c$,
 the cutoff becomes \cite{sudbobrandt,brandt2}
\begin{eqnarray}
S\left( {\bf k} \right)& =& \exp -2g({\bf k}) \nonumber \\
g({\bf k}) & = & \xi_{ab}^2 \left( {\bf k} \times {\bf c} \right)^2 +
 \xi_c^2 \left( {\bf k} \cdot {\bf c} \right)^2 \label{cutoff} \\
& = &\xi_{ab}^2 q^2 + \xi_c^2 ( k^2 - q^2 ) \nonumber
\end{eqnarray}
This provides an elliptical cutoff at large ${\bf k}_\perp =
 (k_x,k_y)$, as required from the shape of the core in Ginzburg-Landau
 theory \cite{klemm}. 
The factor $2$ in the exponential of $S({\bf k})$ is just convention, 
 but it can be determined more accurately by comparison with results 
 from Ginzburg-Landau theory. 
The results in this paper are not affected by  the choice of this 
 parameter.  

The exact form of the cutoff is of some debate.
For straight rigid vortices the form of the cutoff can be derived from
 the shape of the core within Ginzburg-Landau theory\cite{klemm}.
However, the cutoff in Eq. (\ref{cutoff}) depends on $k_z$, the
 component of ${\bf k}$ in direction of the ${\bf B}$ field.
Sardella and Moore\cite{sardellamoore} and Nguyen and
 Sudb\o\cite{nguyensudbo} investigated the tilt-wave instabilities of
 the distorted triangular lattice using a cutoff that depended only on
 ${\bf k}_\perp=(k_x,k_y)$, {\it i.e.}
$g({\bf k})=g({\bf k}_\perp)= \xi_{ab}^2 ({\bf k}_\perp \times {\bf
 c})^2 + \xi_c^2 ({\bf k}_\perp \cdot {\bf c})^2$ in
 Eq. (\ref{cutoff}).
We believe the cutoff in Eq. (\ref{cutoff}) is more physical, as
 it does not depend on being able to specify ${\bf k}_\perp$. That can
 only be done by reference to the average direction of the flux lines,
 {\it i.e.} {\bf B}, but it is hard to believe that the cutoff should
 be sensitive to this overall average direction. 
Some authors do not use a cutoff function $S({\bf k})$, but instead
 introduce an upper limit on any integrations over ${\bf k}$. In most
 situations, if the symmetry of the upper limit introduced is the same
 as the cutoff function $S({\bf k})$, then similar results are
 obtained\cite{nguyensudbo}.
However, in some situations care may be required to ensure it is 
 does not matter whether certain points in  ${\bf k}$-space are 
 just inside or outside the integration range\cite{forgan}.

%%%%%%%%%%%%%%%%%%%%%%%%%%%%%%%%%%%%%%%%%%%%%%%%%%%%%%%%%%%%%%%%%%%%%%
% EQUILIBRIUM LATTICE
%%%%%%%%%%%%%%%%%%%%%%%%%%%%%%%%%%%%%%%%%%%%%%%%%%%%%%%%%%%%%%%%%%%%%%

\section{Equilibrium Lattice}
\label{sec:equillat}

In an isotropic superconductor, the equilibrium flux line lattice 
 is a periodic array where the unit cell is defined by an equilateral 
 triangle. 
Defining the $x$-axis to coincide with one of the basis vectors, the 
 basis vectors may be written as ${\bf R}_1=a\hat{\bf x}$ and 
 ${\bf R}_2=a(\hat{\bf x} + \sqrt{3}\hat{\bf y})/2$ where 
 $a^2 = 2 \Phi_0 / \sqrt{3} B$. 
The presence of anisotropy causes the lattice to distort from this 
 equilateral structure. Using the ideas of anisotropic 
 rescaling\cite{rescale}, the lattice is expected to be of the form
\begin{eqnarray}
{\bf R}_1 &=& a \gamma \hat{\bf x} \nonumber \\
{\bf R}_2 &=& a ( \gamma \hat{\bf x} + \sqrt{3}\hat{\bf y}/ \gamma)/2
\end{eqnarray}
where $\gamma^4 = \cos^2\theta + \epsilon^2 \sin^2\theta $. 
This structure is indeed the equilibrium structure in the limit of 
 the lowest Landau level, and was seen to minimize the free energy 
 of a set of rescaled structures\cite{koganlattice1}. 
All the length scales depend on the strength of the magnetic field 
 in a similar manner i.e. they are proportional to $1/\sqrt{b}$. 
 We refer to  this rescaled structure as the Abrikosov lattice. 

The presence of anisotropy dramatically changes the profile of the 
 magnetic field. 
If the anisotropy is large, the magnetic field associated with a 
 single  isolated flux line contains regions around the flux line 
 where the local magnetic field points in the opposite direction to 
 the average field. 
This allows the usually repulsive flux line  interaction to be 
 attractive. 

Daemen et al.\cite{koganlattice2} showed if we investigate the set of 
 flux line structures with a centered rectangular symmetry, one of 
 which will be the same structure as the Abrikosov lattice, 
 different equilibrium structures can be seen. 
The unit cell consists of an isosceles triangle, see Fig. 2, with two 
 sides of length $l_2$ the other length $l_1$, with an angle $\psi$ 
 between sides of length $l_1$ and $l_2$. The lattice vectors are
\begin{equation}
{\bf R}_{mn} = (m l_1 + n l_2 \cos\psi) \hat{\bf x} + n l_2 \sin\psi 
\hat{\bf y}
\end{equation}
where $m$ and $n$ are integers. 
Repeated computer minimizations of the total free energy have 
 confirmed that for uniaxial anisotropy the assumption the unit cell 
 is an  isosceles triangle is valid\cite{koganlattice2}.  
The magnetic flux per unit cell must be one flux quantum, which 
 allows $l_1$, $l_2$ and $\psi$ to be written in terms of one 
 parameter
\begin{eqnarray}
l_1 &=& \sqrt{ {\Phi_0\over B} {\rho \over \left[ 1 - (\frac12 
 \rho)^2 \right]^{1/2} } } \nonumber\\
l_2 &=& l_1/\rho \\
\cos\psi &=& \frac12 \rho \nonumber 
\end{eqnarray}
Following Daemen et al.\cite{koganlattice2} we minimize the free 
 energy per unit cell $\varepsilon$ using the golden-section-search 
 method\cite{numericalrecipes} because the derivatives of 
 $\varepsilon$ with respect to $\rho$ are hard to calculate. 

The dependence of $l_1$ and $l_2$ on the field, $b=B/H_{c_2}
 (\theta=0)$, is shown in Fig. 3. 
At large fields the two lengths scale as approximately 
 $1/\sqrt{b}$ but there is a smooth crossover to a regime at lower 
 fields where one is approximately constant while the other scales 
 as $1/b$. 
The region where this crossover occurs is characterized by $\kappa$, 
 $\theta$ and $\epsilon$. 
The low field equilibrium state corresponds to the chain state, 
 or `pinstripe structure' observed by Gammel et al.\cite{gammel} 
 in  YBCO. 
This shows care must be taken in defining the flux line lattice in 
 any calculation, as in this low field regime the Abrikosov state is 
 not an  equilibrium state, see Fig. 4. The value of $\rho$ 
 corresponding to the  anisotropically rescaled Abrikosov lattice is 
 given by $\rho_{abr}$. 

The reciprocal lattice of this equilibrium lattice has basis vectors 
\begin{equation}
{\bf Q}_{mn} = n {2\pi\over l_1}\hat{\bf x} + \left[ m 
 {2\pi\over l_2} - n {2\pi\over l_1} \cos\psi \right] 
 {1 \over \sin\psi} \hat{\bf y}
\end{equation}

In the calculations that follow, 
 wavevectors will be measured in units of 
 $( 2\pi/ (3 l_1)\hat{\bf x}, \pi/(l_2 \sin\psi ) \hat{\bf y} ) $, 
 making the rescaled reciprocal lattice vectors $\tilde{\bf Q}_{mn} 
 = 3 n \hat{\bf x} + (2 m - n) \hat{\bf y}$.

%%%%%%%%%%%%%%%%%%%%%%%%%%%%%%%%%%%%%%%%%%%%%%%%%%%%%%%%%%%%%%%%%%%%%%
% ELASTIC INSTABILITIES
%%%%%%%%%%%%%%%%%%%%%%%%%%%%%%%%%%%%%%%%%%%%%%%%%%%%%%%%%%%%%%%%%%%%%%

\section{Elastic Theory of Flux Line Lattice}
\label{sec:inst}

The minimum free energy configuration of the flux lines has been
 assumed to be a periodic array, whose unit cell is an isosceles 
 triangle with the base orientated along the $x$-axis.  
To check the assumption that this free energy is at least a local 
 minimum, the change in the free energy associated
 with small displacements $s_\alpha\left({\bf R}_i(z)\right)$,
 $\alpha=(x,y)$, from the equilibrium lattice ${\bf R}_i = n{\bf R}_1
  + m{\bf R}_2$ can be derived \cite{sudbo1,sardella}. 
Keeping terms only to second order in the displacements $s_\alpha({\bf
 R}_i)$ the change in the free energy is
\begin{equation}
\Delta F = \frac12 \int \hbox{d}^3 {\bf k} s_\alpha\left(-{\bf
 k}\right) \Phi_{\alpha\beta}({\bf k}) s_\beta({\bf k})
\end{equation}
where the integration over ${\bf k}_\perp=(k_x,k_y)$ runs over the
 first Brillouin zone and over $k_z$ on the interval
 $(-\infty,\infty)$. The elasticity matrix is
\begin{equation}
\Phi_{\alpha\beta}\left({\bf k}\right) = {B^2 \over 4\pi}\sum_{\bf Q}
 \left\{ f_{\alpha\beta}\left({\bf k} + {\bf Q}\right) -
 f_{\alpha\beta}\left({\bf Q}\right) \right\} \label{elasticitymatrix}
\end{equation}
where
\begin{eqnarray}
f_{\alpha\beta}\left({\bf p}\right) = p_z^2 V_{\alpha\beta}({\bf p}) 
 &+& p_\alpha p_\beta V_{zz}({\bf p}) - p_z p_\alpha V_{z\beta}
 ({\bf p}) \nonumber \\  & & \qquad - p_z p_\beta V_{z\alpha}({\bf p})
\end{eqnarray}

The stability of a periodic lattice may be determined by
 investigating whether the normal modes of the elasticity
 matrix $\Phi_{\alpha\beta}({\bf k})$ always remain stable {\it i.e.}
 the eigenvalues of $\Phi_{\alpha\beta}({\bf k})$ are positive, or
 whether in some regions of the Brillouin zone the normal modes become
 unstable.

The stability of the distorted triangular lattice (Abrikosov lattice) 
has been examined by various authors
 \cite{sudbobrandt,sardellamoore,nguyensudbo}. 
Sudb\ospace and Brandt\cite{sudbobrandt} observed that for a 
 configuration of rigid flux lines, {\it i.e.} $k_z=0$, as the 
 magnetic field was reduced below a specific level the normal modes 
 became unstable. 
Sardella and Moore\cite{sardellamoore} observed a tilt wave 
 instability, $k_x=k_y=0$, $k_z\neq 0$, and this instability was 
 present in all fields.  
However, the cutoff used did not depend on $k_z$. It was
 suggested that with the use of the cutoff of Eq. (\ref{cutoff}) the 
 instability would disappear\cite{brandt2}.
For ${\bf k}=(0,0,k_z)$ the elasticity matrix is diagonal, and the
 instability calculated by Sardella and Moore was associated with
 $\Phi_{yy}$ becoming negative.
Upon repeating the calculation of Sardella and Moore with the cutoff
 (\ref{cutoff}), we found that the eigenvalue $\Phi_{yy}$ did indeed 
 remain positive.
However, $\Phi_{xx}$ became more unstable, and as in 
 Ref. \cite{sudbobrandt}, the lattice becomes unstable as the field 
 is reduced beyond a critical level. 
Nguyen and Sudb\o\cite{nguyensudbo} have confirmed this by 
 calculating the normal modes of rigid flux lines in the limit $b=0$.

The Abrikosov lattice is only a good approximation of the equilibrium 
 structure at fields $b>O(10^{-2}-10^{-3})$. 
We have investigated the normal modes of the elasticity matrix 
 $\Phi_{\alpha\beta}({\bf k})$ for ${\bf k}=(k_x,k_y,k_z)$, with the 
 equilibrium lattice being the isosceles triangle described in 
 Section \ref{sec:equillat}, using the $k_z$ dependent 
 cutoff (\ref{cutoff}). 
The presence of instabilities is expected as Sardella\cite{sardella2}
  has  confirmed the existence of a zone center instability for 
 large tilt angles $\theta$, large anisotropy $1/\epsilon$ and small 
 fields $b$. 

The fields at which instabilities are observed depend on $\kappa$, on 
 the tilt angle $\theta$ and on the anisotropy $\epsilon$. 
The isotropic system, $\epsilon=1$, is stable. 
As the anisotropy increases, $\epsilon$ decreases, no instabilities 
 are observed until a critical anisotropy is obtained. 
The minimum anisotropy required for instabilities to be seen depends 
 on $\kappa$. For $\kappa=20$ instabilities were seen for 
 $1/\epsilon^2 > 120.3$, but this was increased to $1/\epsilon^2 > 
 138$ for $\kappa=50$. For values of $1/\epsilon$ less than these 
 critical values the lattice is stable at {\it all} orientations. 

If the anisotropy is larger that this critical anisotropy, 
 instabilities can be seen. 
The lattice is always stable at large fields. 
As the field is reduced there is a specific field 
 $b^*(\kappa,\epsilon,\theta)$ at which the lattice {\it initially} 
 becomes unstable. 
Fig. 5 shows that for a given angle $\theta$ there is a minimum 
 anisotropy $1/\epsilon$ below which the lattice is stable in all 
 fields. It also shows that the critical field $b^*$ increases as 
 the system becomes more anisotropic. 
The lattice is always stable at $\theta=0$ and $\theta=\pi/2$, but 
 once the anisotropy is large enough there exists a range of angles 
 at which instabilities are seen, see Fig. 6. This range of angle 
 increases as the anisotropy increases. 

The wavevectors at which these instabilities are first seen, 
 ${\bf k}^*(\kappa,\epsilon,\theta)$ always have {\it finite} $k_z$. 
Both $b^*$ and ${\bf k}^*$ are functions of $\kappa$, $\epsilon$ and 
 $\theta$. 
These unstable modes correspond to displacements of the lattice
 approximately parallel to the $x$-axis, and we call it the
 `staircase wave' instability to distinguish it from other 
 instabilities observed. 
In Fig. 7 the dependence of the $x$ and $y$ components 
 of $k^*$ are shown; $k_x$ and $k_y$ have been rescaled so that the 
 dashed line shows the edge of the first Brillouin zone. $\theta$ 
 decreases from small $k_x$ to large $k_x$. 
The explanation of the behavior of the actual wavevector where the
 instability first appears, ${\bf k}^*$, is not obvious to us.

The parameters chosen in this paper ($\epsilon=1/60$, $\kappa=50$)
 were chosen to give some insight into the behavior of BSCCO.
While the exact value of the anisotropy in BSCCO is unknown, the
 values used in this paper are similar to those used in other
 papers\cite{sudbobrandt}\cite{sardellamoore}\cite{nguyensudbo}.
In comparison, YBCO is much less anisotropic, with $\epsilon \approx 
 1/5$ and $\kappa \approx 50$, and as can be seen from Fig. 5 is 
 well below that required for the instabilities to be observed. 
To explore the instabilities in full detail as a function of $\kappa$,
 $\epsilon$, $\theta$ and $b$ is very time consuming, but Fig. 5 
 shows the `chain' state is stable for parameters that could 
 describe YBCO, but unstable for BSCCO.

%%%%%%%%%%%%%%%%%%%%%%%%%%%%%%%%%%%%%%%%%%%%%%%%%%%%%%%%%%%%%%%%%%%%%%
% H_{c_1}: Abrikosov Limit
%%%%%%%%%%%%%%%%%%%%%%%%%%%%%%%%%%%%%%%%%%%%%%%%%%%%%%%%%%%%%%%%%%%%%%

\section{Lower Critical Field: Abrikosov Lattice Limit}
\label{sec:angle}

The presence of elastic instabilities for very anisotropic materials 
 may indicate why there are differences in the Bitter patterns 
 observed for BSCCO and YBCO. 
One interpretation of the BSCCO patterns was the coexistence of two 
 interpenetrating lattices\cite{huse,sudbobrandthuse}. This 
 postulate has been investigated by calculating the angles at which 
 single flux lines first enter a superconducting 
 sample\cite{sudbobrandt,nguyensudbo}. 
We modify the calculation of Nguyen and Sudb\o\cite{nguyensudbo} to 
 show this possible effect in order to emphasize the importance of 
 the cutoff procedure used.

We consider a cylindrical superconducting sample, with the applied
 field ${\cal H}$ perpendicular to the axis of the cylinder, and
 tilted at an angle $\phi$ away from the crystal c-axis, see Fig. 8. 
This geometry is chosen so that demagnetization effects permit 
 solutions where the flux lines are straight for all orientations 
 of the applied field. 

Within this geometry, the Gibbs free energy for a system of rigid
 straight flux lines, tilted at (a different) angle $\theta$ is
\begin{equation}
G=F-{B {\cal H} \cos(\phi-\theta)\over 4\pi}
\end{equation}
Neglecting the interaction between the flux lines, the free energy 
 within the London approximation (\ref{londonfree2}) is
\begin{equation}
F= {B^2\over 8\pi} \int { \hbox{d}^2{\bf q} \over 4\pi^2 } S({\bf q}) 
 {1 + \lambda_\theta^2 q^2 \over ( 1 + \lambda_{ab}^2 q^2) 
 (1 + \lambda_{\theta}^2 q_x^2 + \lambda_c^2 q_y^2) }
\end{equation}
where $\lambda_{\theta}^2 = \lambda_{ab}^2 \sin^2\theta +
 \lambda_c^2\cos^2\theta$. As with most calculations within the London
 approximation, the integral is formally divergent without the cutoff 
 term, but the use of the cutoff $S({\bf q})$ described in the 
 previous section allows the calculation to proceed. 
Also, as  the integration 
 is over ${\bf q}$ perpendicular to the flux line this calculation is 
 insensitive to the $q_z$ dependence of the cutoff $S({\bf q})$. 
 By neglecting the interaction 
 between the flux lines we are essentially just discussing the 
 behavior  of a single flux line. This calculation can be viewed as 
 the extension of the Abrikosov lattice to $b=0$.

As the field ${\cal H}$ is increased, the flux lines initially 
 enter the sample when $G=0$. 
Although it would be preferable to chose $\phi$, the orientation of 
 the applied field, and then calculate $\theta$ and ${\cal H}$ it 
 is only possible to assume $\theta$ and then calculate the 
 corresponding values of ${\cal H}$ and $\phi$. 
Looking for solutions where $\pt G /\pt\theta = 0$ implies the 
 orientation(s) of the flux lines within the sample are governed by 
 the relation
\begin{equation}
\tan\phi = { \tan\theta + F^\prime/F \over 1 - F^\prime\tan\theta /F} 
\label{phitheta}
\end{equation}
When investigating the first entry of flux lines into the sample, it
 is well known that the cutoff used is important. If a circular cutoff
 is used then there is a unique orientation of the flux lines, given
 by\cite{sudbobrandthuse,nguyensudbo}
\begin{equation}
\tan\phi = \epsilon^{2} \tan\theta
\end{equation}
However, this is not true if an elliptical cutoff is used, as for 
 large anisotropies there exists a nonmonotonic relation between 
 $\theta$ and $\phi$, see Fig. 9. This implies the 
 possibility of the existence of two orientations of flux lines. 

Within the nonmonotonic regime, there exist three possible
 orientations of the flux lines, $\theta_a$, $\theta_b$, and
 $\theta_c$, for any given $\phi$. $\theta_b$ corresponds to an
 unstable orientation of the flux lines {\it i.e. }a maximum in the
 Gibbs free energy, and will not be considered further.

By calculating the Gibbs free energy, we can see whether the flux
 lines will orientate at either $\theta_a$ or $\theta_c$, and the
 relationship between $\phi$ and $\theta$ determined.
A `forbidden' region occurs due to the presence of the nonmonotonic
 regime. The flux line lattice can only be orientated at angles
 $\theta < \theta^*_1$ and $\theta> \theta_2^*$.

This can be seen by noting the relationship between the calculated 
 values of ${\cal H}$ and $\phi$. Fig. 10 shows ${\cal H}(\phi)$ when 
 all flux line orientations are allowed. 
The points in Fig. 10 are equally spaced in $\theta$. 
For large anisotropy ${\cal H}(\phi)$ becomes nonmonotonic when 
 $\theta(\phi)$ becomes nonmonotonic. 
In Fig. 11 the three values of ${\cal H}(\phi)$ in the 
 nonmonotonic regime correspond to the three possible orientations 
 of the flux lines at $\theta_a$, $\theta_b$ and $\theta_c$. 
While all three orientations correspond to $G=0$ and 
 $\pt G/\pt \theta=0$, only one is a global minimum in $G(\theta)$. 
This global minimum will be  the equilibrium structure and is the 
 orientation with the smallest value of ${\cal H}(\phi)$. 
The lower critical field $H_{c_1}(\phi)$ is therefore just the 
 smallest value of ${\cal H}(\phi)$ for any given $\phi$. 
From Fig. 11 it can be seen $H_{c_1}$ has a kink where 
 ${\cal H}(\theta_1^*) = {\cal H}(\theta_2^*)$ and 
 $\phi(\theta_1^*) = \phi(\theta_2^*)=\phi_{kink}$. 
The presence of kinks in the lower critical field for single flux 
 lines have also been observed within the Ginzburg-Landau 
 model\cite{klemm}.

If we now allow the superposition of non-interacting flux lines, 
 which are assumed to be far apart in the limit $b\to 0$, the 
 `forbidden region' may be removed. 
At $H_{c_1}(\phi_{kink})$ the average field ${\bf B}$ may be 
 orientated at all angles within the `forbidden' region 
 $\theta_1^*<\theta<\theta_2^*$ by orientating some flux lines at 
 $\theta_1^*$ and the others at $\theta_2^*$.  

%CUTOFF

Various authors have discussed the competition between the 
 coexistence of different flux line species and the presence of 
 elastic instabilities in the supposed equilibrium structure. 
For single isolated flux lines (in the limit $b\to 0$) the elastic 
 instabilities depend most strongly on $k_z$ and only tilt-wave 
 instabilities will be discussed further in this section. 

As $k_z$ is non-zero, the cutoff used is important. 
 Nguyen and Sudb\ospace used $S({\bf k})=S({\bf k}_\perp)$. For any 
 given value of $\kappa$ they found two important values of the 
 anisotropy. At small anisotropies the relationship between $\phi$ 
 and $\theta$ is monotonic. 
For $1/\epsilon > \Gamma_1^{NS} $ the relationship between $\phi$ and 
 $\theta$ becomes nonmonotonic and the Gibbs free energy may be 
 doubly degenerate. 
However, if $1/\epsilon > \Gamma_2^{NS}$ the `perpendicular' tilt 
 modulus $c_{44}^\perp ({\bf k}) = \Phi_{xx}(0,0,k_z)/k_z^2$ becomes 
 negative over a range of angles. The two critical anisotropies 
 depend linearly on $\ln \kappa$, but $\Gamma_2^{NS}$ is much larger 
 than $\Gamma_1^{NS}$. 

There is a marked difference if the cutoff 
 $S({\bf k})=\exp(-2g({\bf k}))$, $g({\bf k}) = \xi_{ab}^2 
 ({\bf k}\times {\bf c})^2 + \xi_c^2 ({\bf k} \cdot {\bf c})^2 $. 
The dependence of $\Gamma_1^{NS}$ and $\Gamma_2^{NS}$ on 
 $\kappa$ is similar, but $\Gamma_1^{NS} = \Gamma_2^{NS} = \Gamma$. 
For $\kappa=50$ and $S({\bf k})=\exp(-2g({\bf k}))$ it is found 
 $\Gamma \approx 9.65$. The value of $\Gamma$ does depend on the 
 cutoff procedure used. Using a cutoff of the same symmetry but a 
 different strength, $S({\bf k})=\exp(-g({\bf k}))$, it is found 
 $\Gamma \approx 10.03$ for $\kappa=50$

These two `competing' effects are both related to the line tension 
 $P_l(\theta) = \varepsilon_l(\theta) + \pt^2 \varepsilon_l(\theta) / 
 \pt\theta^2$, where $\varepsilon$ is the line energy. 
In the limit $b\to 0$ where the London free energy 
 $F=n\varepsilon_l$, $n$ is the areal density of flux lines, then 
 it follows from (\ref{phitheta}) that
\begin{equation}
P_l(\theta) = \varepsilon_l(\theta) \sec^2(\phi-\theta){\pt \phi 
 \over \pt \theta}
\end{equation}
However, $c_{44}$ may be defined as 
 $c_{44}= d^2 F(\theta)/d\theta^2$. For the isolated noninteracting 
 flux lines $F= n\varepsilon_l$, so  $c_{44} \propto \varepsilon_l
 (\theta) + \pt^2 \varepsilon_l(\theta) / \pt\theta^2 = 
 P_l(\theta)$\cite{sudbobrandt2}, where the first term is 
 due to the compression of the flux line lattice during tilting. 
Therefore the presence of a kink in $H_{c_1}(\phi)$ not only causes a 
 restriction of orientations of the flux line lattice, and the 
 possibility of the coexistence of different flux line orientations, 
 but is also related to the presence of a tilt-wave instability. 
The tilt-wave instabilities only occur where $\pt \phi / \pt \theta 
 < 0$, see also Grishin et al.\cite{grishin} , but this is only part 
 of the region of excluded flux line orientations $\theta_1^* < 
 \theta < \theta_2^*$ and the instabilities are never observed.

%%%%%%%%%%%%%%%%%%%%%%%%%%%%%%%%%%%%%%%%%%%%%%%%%%%%%%%%%%%%%%%%%%%%%%
% H_{c_1}: Chain State
%%%%%%%%%%%%%%%%%%%%%%%%%%%%%%%%%%%%%%%%%%%%%%%%%%%%%%%%%%%%%%%%%%%%%%

\section{Lower Critical Field: Chain State Limit}
\label{sec:chain}

The previous calculation assumed the equilibrium low field flux line 
 structure was a configuration of well separated non-interacting flux 
 lines ,{\it i.e.} the extension of the Abrikosov lattice to $b=0$. 
In section \ref{sec:equillat} it was seen the low field equilibrium 
 structures are very different from the Abrikosov lattice. 
The anisotropy induced attractive interaction makes it easier for 
 chains of flux lines to enter a sample than for single flux lines, 
 which forces $H_{c_1}$ for an single infinite chain to be less 
 than $H_{c_1}$ for a  single flux line\cite{grishin}. 
We take the limit $b\to 0$ to imply there is only one vortex chain 
 in the sample, the separation between the chains being effectively 
 infinite. 

The presence of instabilities in the `chain' state were investigated 
 in  Section \ref{sec:inst}, which seemed to indicate a different 
 behavior for materials such as YBCO from those such as BSCCO. 
These instabilities were seen as the field was reduced from a stable 
 regime similar to the Abrikosov lattice. In this section we 
 investigate what happens as the flux lines initially enter the 
 sample {\it i.e.} $b=0$. 
Chosing the same geometry as section \ref{sec:angle} we again look 
 at the  relationship between $\phi$ and $\theta$ and relate this 
 to the elastic  instabilities. 

For a single flux line in section \ref{sec:angle}, the 2D line energy 
 integral is easily calculated by rescaling the coordinates, see 
 Nguyen and Sudb\ospace\cite{nguyensudbo}. 
For the single infinite chain, this 2D integral is replaced by a sum 
 over the reciprocal lattice vectors of the 1d chain and a 1d 
 integral perpendicular to the chain. 
Assuming the chain to be a set of flux lines equally spaced along the 
 $x$-axis, then 
\begin{equation}
\int { \hbox{d}^2{\bf q} \over (2\pi)^2 } \to {1 \over l_{ch}} \sum_n 
 \int { \hbox{d} q_y \over 2\pi } 
\end{equation}
where $q_x = 2\pi n/l_{ch}$, $n=0 \pm 1 \pm 2...$ , $l_{ch}$ being 
 the separation of the flux lines in the chain. $l_{ch}$ is 
 determined by finding the minimum in the line energy, and is a 
 nonmonotonic function of $\theta$.

The behavior of the chain state is subtly different from that of the 
 single flux line. The main effects are the same, {\it i.e.} for 
 large anisotropy  there exist a range of angles over which the 
 chains cannot be orientated, but the details are more complicated. 

The chain state is stable for small anisotropies, but for large 
 anisotropies the relationship between $\phi$ and $\theta$ once again 
 becomes nonmonotonic. 
For this to occur it requires a larger anisotropy than for the single 
 flux line. Fig. 12 shows $\phi(\theta)$ just after the onset of 
 nonmonotonicity for isolated flux lines. 
The chain state has no restriction on the orientation of the flux 
 lines, unlike the single flux line, and also the chain state always 
 has a lower (or equal) lower critical field 
 $H_{c_1}$\cite{grishin}, see Fig. 13. 

Increasing the anisotropy, Fig. 14, the chain state eventually shows 
 signs of instabilities. 
Initially $\phi(\theta)$ becomes nonmonotonic at large tilt angles, 
 $\theta\approx 9\pi/20$, and the chains are excluded over the range 
 $\theta_{2^\prime}^* < \theta < \theta_2^*$. This occurs for 
 $1/\epsilon > \Gamma_1(\kappa) $ and $H_{c_1}(\phi)$ contains a 
 kink in a manner similar to the single flux line.  

However, for $1/\epsilon > \Gamma_2(\kappa)$, $\phi(\theta)$ is also 
 nonmonotonic at smaller angles, $\theta\approx\pi/4$. This implies 
 the chains are also excluded over a different range 
 $\theta_1^* < \theta < \theta_{1^\prime}^*$. 
From ${\cal H}(\phi)$, Fig. 15,  it can be seen that a stable range 
 $\theta_{1^\prime}^* < \theta < \theta_{2^\prime}^*$ exists between 
 these two forbidden regions. 
The lower critical field $H_{c_1}(\phi)$ now  has two kinks. 
As the anisotropy is increased further these excluded regions grow 
 until at $1/\epsilon > \Gamma_3(\kappa)$ the chain state is 
 excluded over the whole range $\theta_1^* < \theta < \theta_2^*$. 
 The second kink in $H_{c_1}(\phi)$ disappears, Fig. 16, and only 
 one kink remains. For $\kappa=50$, $\Gamma_1 \approx 15$, 
 $\Gamma_2 \approx 30$ and $\Gamma_3 \approx 60$. 

The elastic instabilities again only occur when $\pt \phi /\pt \theta 
 < 0$. 
These instabilities have slightly different properties at small 
 $\theta$  and large $\theta$. 
At large $\theta$ the lattice is most unstable to a tilt-wave 
 instability ${\bf k}=(0,0,k_z)$ while it is a staircase wave 
 instability, with $k_x$ and $k_z$ both non-zero, that causes the 
 chain to become unstable at smaller $\theta$. 
Again, these instabilities only occur in the forbidden regions, and 
 the chain state is stable at all allowed orientations. 

%%%%%%%%%%%%%%%%%%%%%%%%%%%%%%%%%%%%%%%%%%%%%%%%%%%%%%%%%%%%%%%%%%%%%%
% Conclusion
%%%%%%%%%%%%%%%%%%%%%%%%%%%%%%%%%%%%%%%%%%%%%%%%%%%%%%%%%%%%%%%%%%%%%%

\section{Conclusion}

We have investigated the presence of instabilities in the flux line
 lattice in anisotropic superconductors using London theory.
There are limitations on the applicability of the London theory, but
 the wavevectors at which the instabilities appear are well within its
 limits of validity, {\it i.e.} ${\bf k} \ll O(1/\xi)$.

Due to the anisotropy induced attractive flux line interaction, the 
 form of the equilibrium flux line lattice has to be determined 
 numerically at all  fields. 
There is a smooth crossover from a large field Abrikosov-like state 
 to the  low field chain-like state. The nature of the instabilities 
 observed in this equilibrium lattice is not complex. 
At large fields it is stable. 
As the field is reduced, at a particular field $b^*$ some of the
 normal modes of the elasticity matrix become unstable. This field
 depends on the anisotropy $\epsilon$ and the angle $\theta$ at which
 the flux lines are orientated to the crystal $c$-axis. 
The instability is characterized by always having finite $k_x$ and 
 $k_z$, and we call this a staircase wave instability to distinguish 
 it from a tilt wave instability which is defined as an instability 
 which depends only on $k_z$. 
There is a minimum anisotropy required before these instabilities are 
 observed. This depends on $\kappa$ and for $\kappa=20$ the 
 instabilities were seen for $1/\epsilon^2 > 120.3$, but this was 
 increased to $1/\epsilon^2 > 138$ for $\kappa=50$. 
This may indicate why the Bitter patterns observed on YBCO are 
 different from those on BSCCO. 

As the component of ${\bf k}$ along the flux lines is non-zero, the 
 choice of cutoff is crucial. We have used the cutoff suggested by 
 Sudb\ospace and Brandt, where 
 $ S\left( {\bf k} \right) = \exp -2g({\bf k})$ and 
 $ g({\bf k})  =  \xi_{ab}^2 \left( {\bf k} \times {\bf c} \right)^2 
 + \xi_c^2 \left( {\bf k} \cdot {\bf c} \right)^2 $. 
This cutoff appears more physical than a cutoff that just depends on 
 ${\bf k}_\perp$ and allows the elastic instabilities and the 
 nonmonotonic behavior of $\phi(\theta)$ to be related to a single 
 quantity, the line tension $P_l(\theta)$, as expected. 
However, it should be noted there are still problems associated with
 this cutoff in the limit $k_z \to \infty$. Once in the region
 where $k_z$ forces $S\left( {\bf k} \right) \ll 1$, the elasticity
 matrix (\ref{elasticitymatrix}) will be dominated by $-\sum_{\bf
 Q}f_{\alpha\beta}\left({\bf Q}\right)$.
While $\Phi_{xy}$ is zero, the eigenvalues of the elasticity matrix
 are negative, showing London theory is always unstable at all fields
 and angles.
However, these instabilities occur at values of $k_z \approx O(1/\xi)$
 and are unphysical.  
The inclusion of core effects, {\it e.g.} core bending energy, may 
 remove remove this instability, but this is outside the domain of
 validity of the London approximation. 

Instabilities can also be observed by investigating the lower 
 critical field $H_{c_1}$, {\it i.e. } $b=0$. 
These $b=0$ instabilities are different from the large field 
 staircase wave  instabilities. 
When the applied field ${\cal H}$ is tilted away from the 
 crystal axis, the ${\bf B}$-field and the applied field ${\cal H}$ 
 are not parallel. 
In Section \ref{sec:angle} it was seen that for small fields and
 large anisotropies there existed a nonmonotonic relationship between
 the angle $\theta$ at which the ${\bf B}$-field is tilted away from
 the $c$-axis, and the angle $\phi$ at which the applied field is
 tilted. 
The nonmonotonicity is related to the elastic instabilities 
 observed, but also imply there is a restriction on the allowed 
 orientations of the ${\bf B}$-field and a kink in $H_{c_1}(\phi)$. 
The properties of the chain state changes at three different values 
 of the  anisotropy. For $1/\epsilon < \Gamma_1$, all possible 
 orientations of the chain state are possible, while in the interval 
 $\Gamma_1 < 1/\epsilon < \Gamma_2$ the chains cannot be orientated 
 over the range $\theta_{2^\prime}^* < \theta < \theta_2^*$. 
However, if $\Gamma_2 < 1/\epsilon < \Gamma_3 $ the flux lines are 
 excluded over two separate regions which grow to one large region 
 for $1/\epsilon > \Gamma_3$. This unusual behavior can also be 
 described as $H_{c_1}(\phi)$ developing one kink at $1/\epsilon = 
 \Gamma_1$, a second kink developing at $1/\epsilon=\Gamma_2$ but 
 $H_{c_1}(\phi)$ only having a single kink for $1/\epsilon > 
 \Gamma_3$.

Whether the peculiar effect of having the allowed orientations 
 $\theta_{1^\prime}^* < \theta < \theta_{2^\prime}^*$ between two 
 forbidden regions really exists is unclear. 
Just as the existence of the forbidden region  
 $\theta_1^* < \theta < \theta^*_2$ for the isolated flux lines in 
 Section \ref{sec:angle} may have been 
 an indicator of a state with a lower free energy and applied field 
 $H_{c_1}$, there may exist a new composition of flux lines that will 
 be stable at large anisotropies and have a lower free energy and 
 applied field $H_{c_1}$ than the chain state. 
This may be a completely new state or a superposition of chain 
 states and other flux lines, but a full investigation of such 
 states is left for future work.

\acknowledgements{This work was funded by EPSRC grant GR/J60681.}

%%%%%%%%%%%%%%%%%%%%%%%%%%%%%%%%%%%%%%%%%%%%%%%%%%%%%%%%%%%%%%%%%%%%%%
% REFERENCES
%%%%%%%%%%%%%%%%%%%%%%%%%%%%%%%%%%%%%%%%%%%%%%%%%%%%%%%%%%%%%%%%%%%%%%

%%%%%%%%%%%%%%%%%%%%%%%%%%%%%%%%%%%%%%%%%%%%%%%%%%%%%%%%%%%%%%%%%%%%%%
% FIGURES
%%%%%%%%%%%%%%%%%%%%%%%%%%%%%%%%%%%%%%%%%%%%%%%%%%%%%%%%%%%%%%%%%%%%%%
%\section{Figures}

\begin{figure}
  \narrowtext
  \centerline{\epsfxsize=6.0cm
  \epsfbox{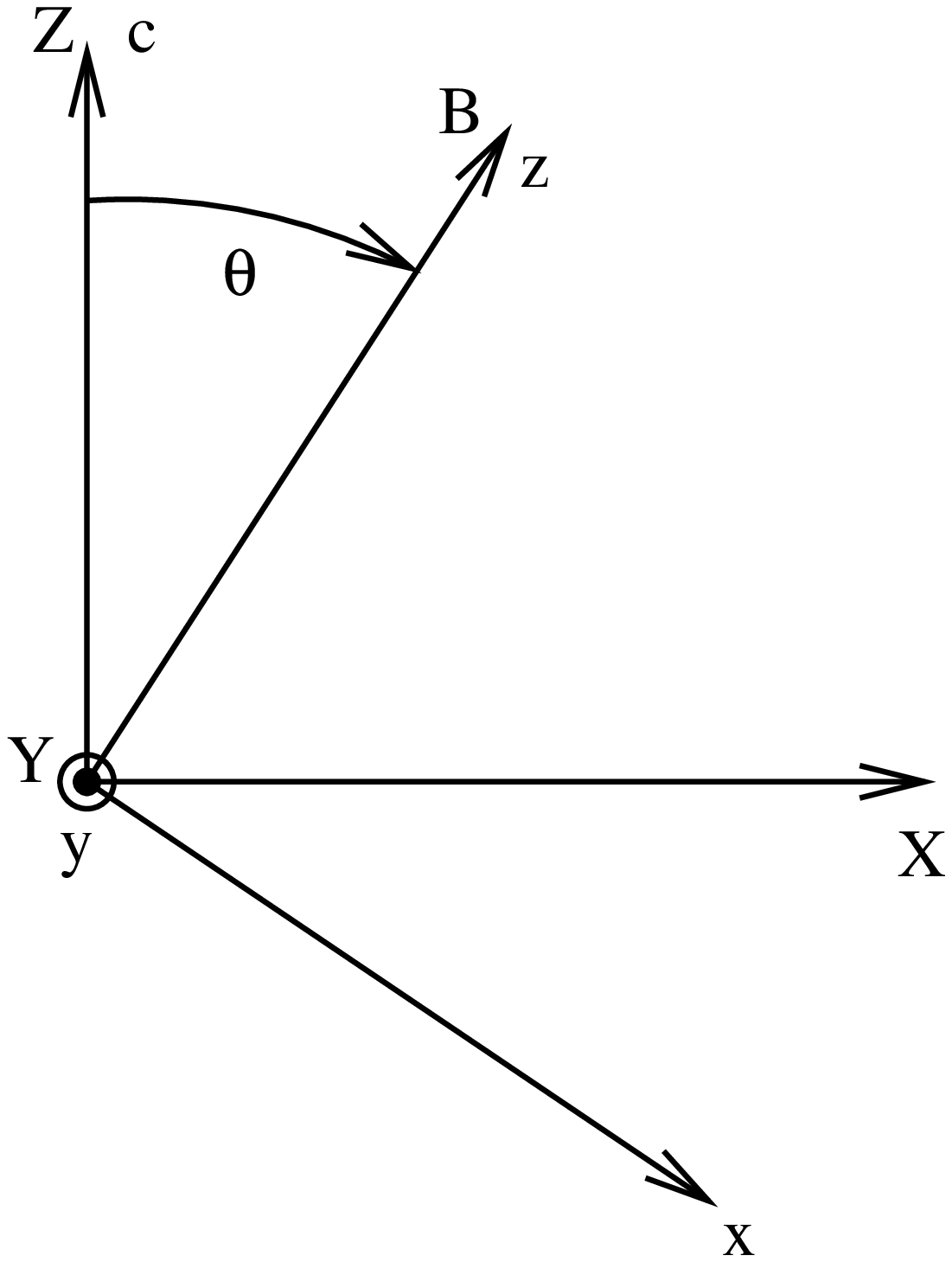} }
  \caption{The `vortex' frame $(x,y,z)$ is obtained by rotating the 
   crystal frame $(X,Y,Z)$ by an angle $\theta$ about the $Y$ axis.}
\end{figure}

\vglue 1in

\begin{figure}
  \narrowtext
  \centerline{\epsfxsize=6.0cm
  \epsfbox{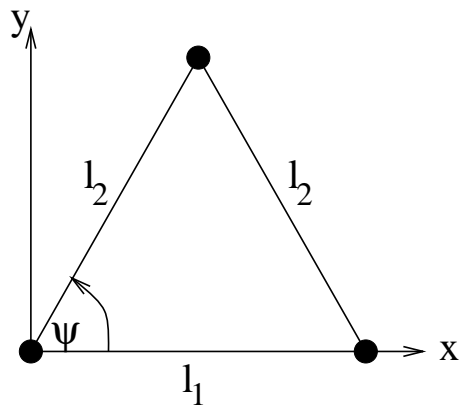} }
  \caption{The unit cell of `chain' state.}
\end{figure}
     
\newpage
\begin{figure}
  \narrowtext
  \centerline{\epsfxsize=6.5cm
  \epsfbox{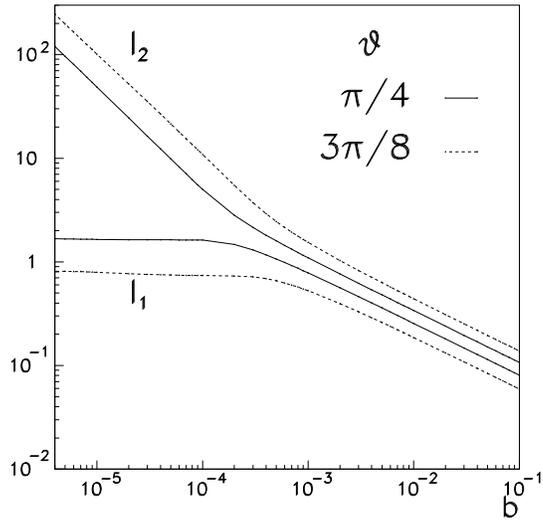} }
  \caption{The field dependence of the separation of the flux lines 
  (measured in $\lambda_{ab}$) for $\epsilon=1/60$ and $\kappa=50$. 
  At large fields $l_1$ and $l_2$ are proportional to 
  $1/\protect{\sqrt{b}}$ but at lower fields there is a crossover to 
  the `chain' state.}
\end{figure}
 
\begin{figure}
  \narrowtext
  \centerline{\epsfxsize=6.0cm
  \epsfbox{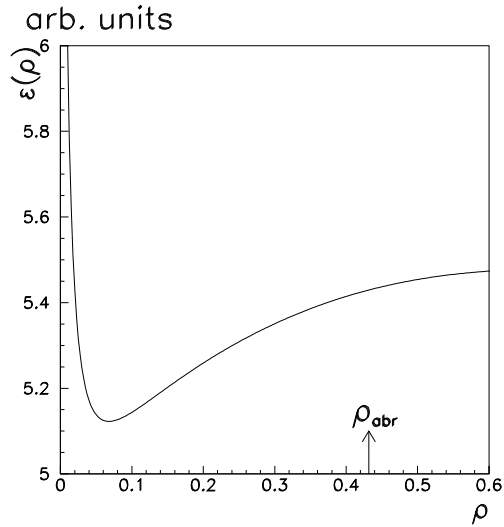} }
  \caption{The dependence of the energy per flux line on the parameter 
   $\rho$, for $\epsilon=1/60$, $\theta=3\pi/8$, $\kappa=50$, 
   $b=10^{-4}$. The arrow marks the value of $\rho$ of the rescaled 
   Abrikosov lattice. 
   This figure  clearly shows that the Abrikosov lattice is not the 
   solution of minimum free energy.}
\end{figure}

\newpage
\begin{figure}
  \narrowtext
  \centerline{\epsfxsize=6.5cm
  \epsfbox{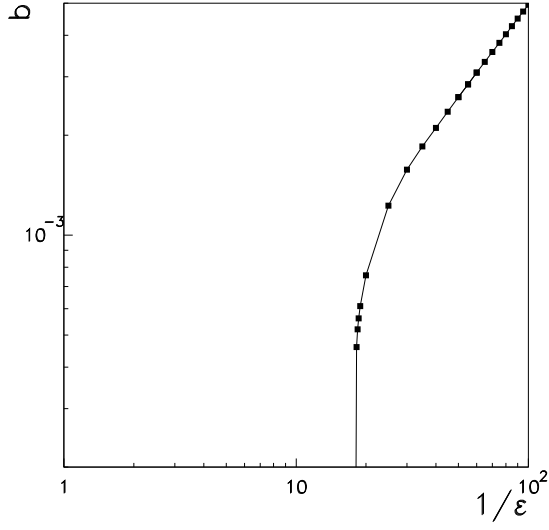} }
  \caption{The field at which the lattice initially becomes unstable 
  as a function of the anisotropy $\epsilon$, for $\theta=3\pi/8$ and 
  $\kappa=50$.}
\end{figure}

\begin{figure}
  \narrowtext
  \centerline{\epsfxsize=6.5cm
  \epsfbox{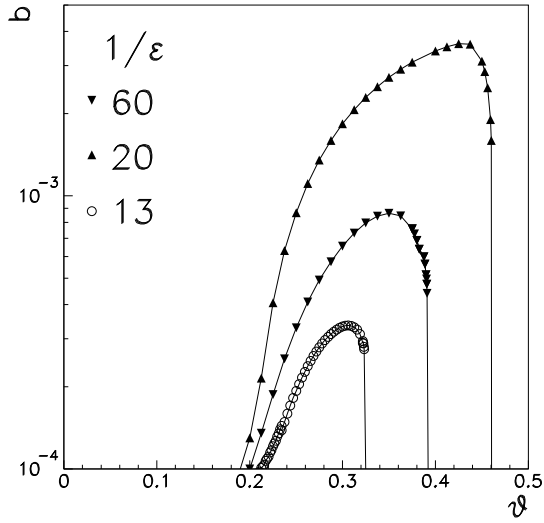} }
  \caption{The field at which the lattice initially becomes unstable as a 
  function of the angle, $\theta$, by which the lattice is tilted from 
  the $c$-axis, for different values of $\epsilon$ and $\kappa=50$.}
\end{figure}

\newpage
\begin{figure}
  \narrowtext
  \centerline{\epsfxsize=6.5cm
  \epsfbox{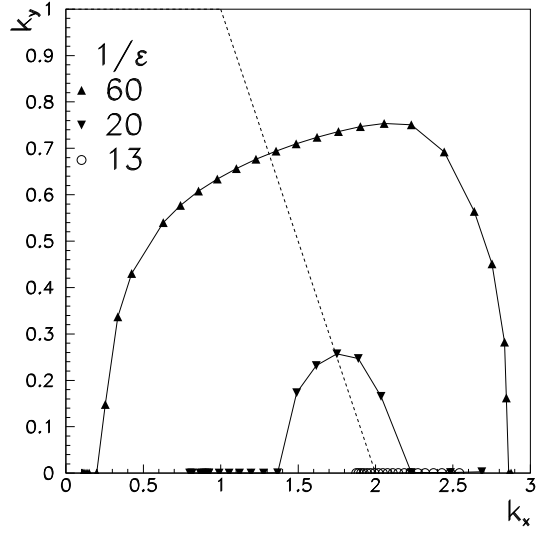} }
  \caption{The rescaled $k_x$ and $k_y$ components of ${\bf k}^*$, 
  for different $\epsilon$ and $\kappa=50$. The dashed line marks 
 the edge of the first Brillouin zone.}
\end{figure}

\begin{figure}
  \narrowtext
  \centerline{\epsfxsize=6.5cm
  \epsfbox{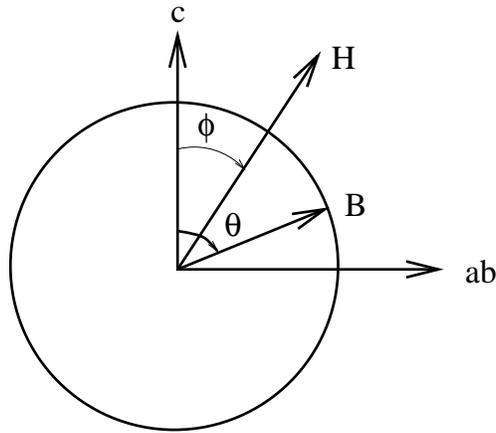} }
  \caption{The geometry of sample and applied field ${\cal H}$. 
  The applied field is tilted at an angle $\phi$ from the c-axis, while 
  the magnetic induction ${\bf B}$ is tilted at an angle $\theta$.}
\end{figure}

\newpage
\begin{figure}
  \narrowtext
  \centerline{\epsfxsize=6.5cm
  \epsfbox{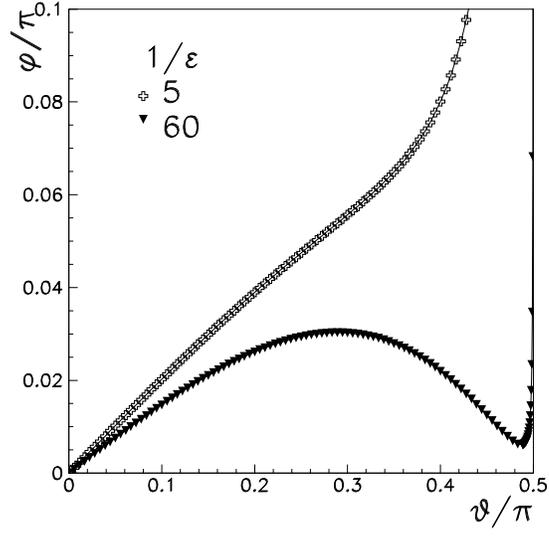} }
  \caption{The relationship between $\theta$ and $\phi$ for 
  different anisotropies, with $\kappa=50$. This relationship becomes 
  nonmonotic over a range of angles for large anisotropy.}
\end{figure}

\begin{figure}
  \narrowtext
  \centerline{\epsfxsize=6.5cm
  \epsfbox{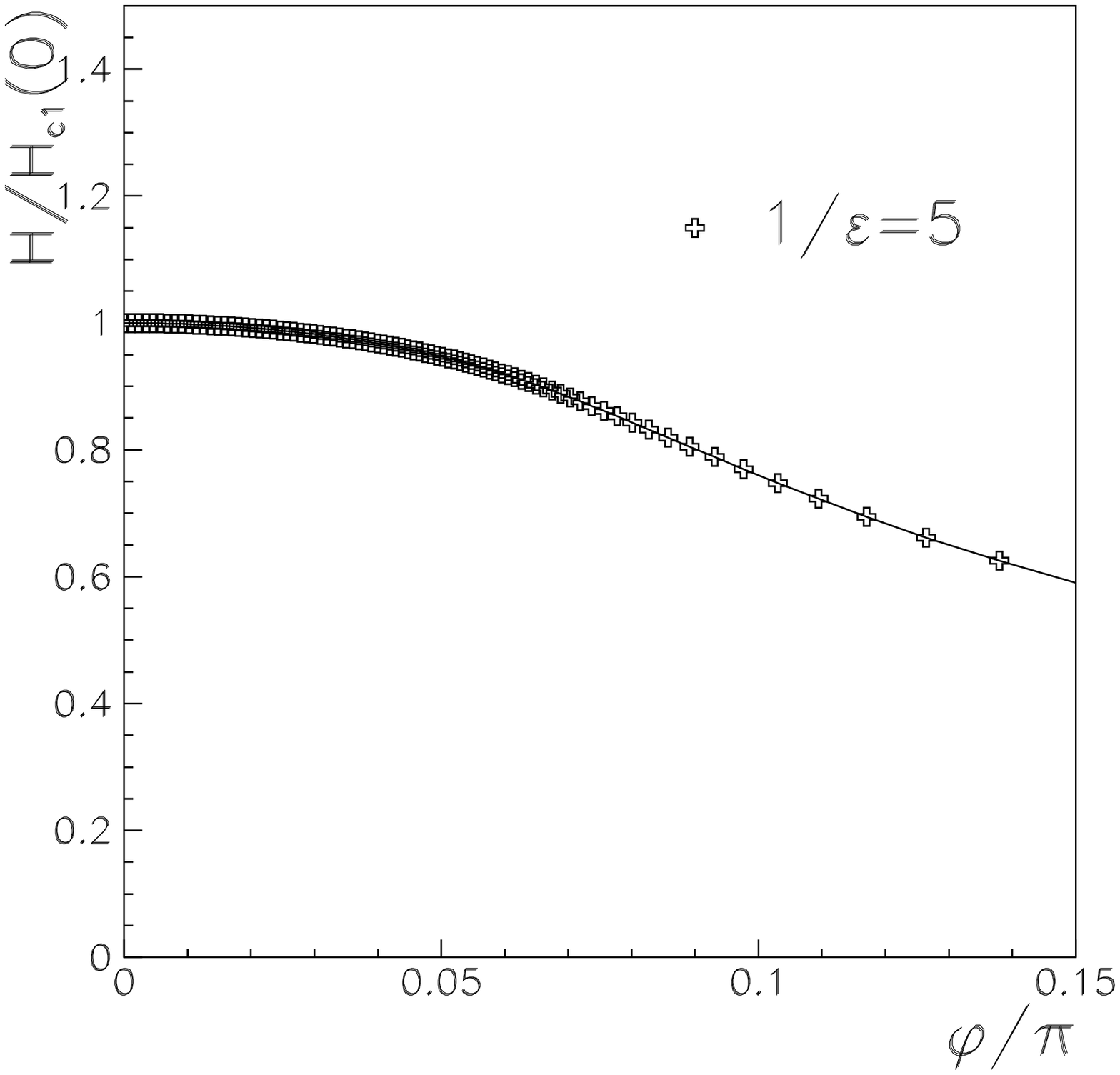} }
  \caption{The applied field ${\cal H}(\phi)$ for $\epsilon=1/5$ and 
  $\kappa=50$.}
\end{figure}
 
\newpage
\begin{figure}
  \narrowtext
  \centerline{\epsfxsize=6.5cm
  \epsfbox{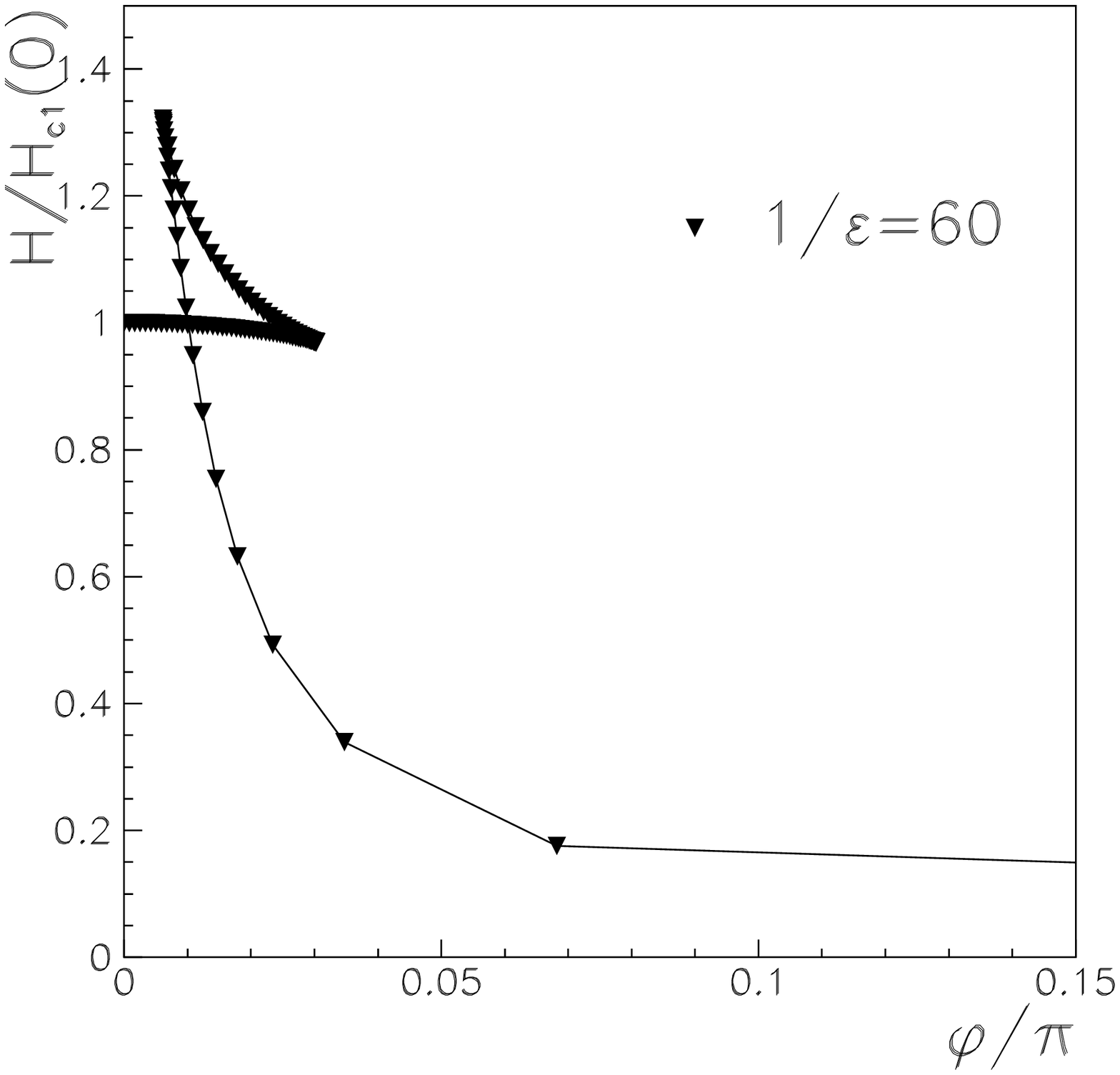} }
  \caption{The applied field ${\cal H}(\phi)$ for $\epsilon=1/60$ and 
   $\kappa=50$.}
\end{figure}

\begin{figure}
  \narrowtext
  \centerline{\epsfxsize=6.5cm
  \epsfbox{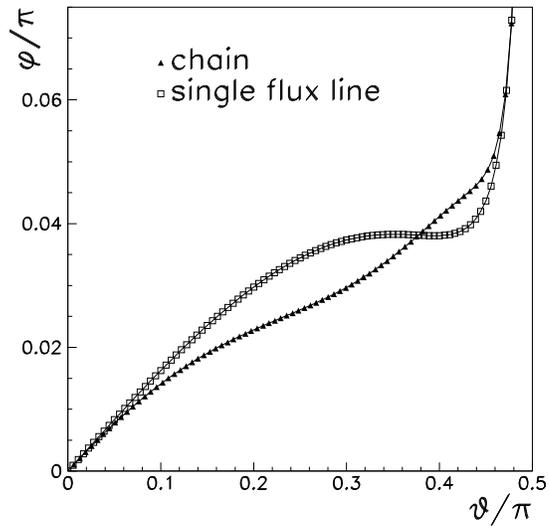} }
  \caption{The relationship between $\phi$ and $\theta$ for 
   different lattices, with $\kappa=50$ and $\epsilon=1/10$.}
\end{figure}

\newpage
\begin{figure}
  \narrowtext
  \centerline{\epsfxsize=6.5cm
  \epsfbox{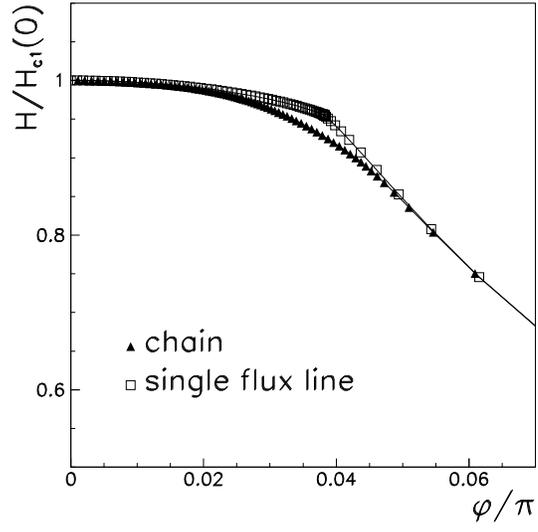} }
  \caption{The applied field ${\cal H}(\phi)$ for different lattices, 
   with $\kappa=50$ and $\epsilon=1/10$.}
\end{figure}

\begin{figure}
  \narrowtext
  \centerline{\epsfxsize=6.5cm
  \epsfbox{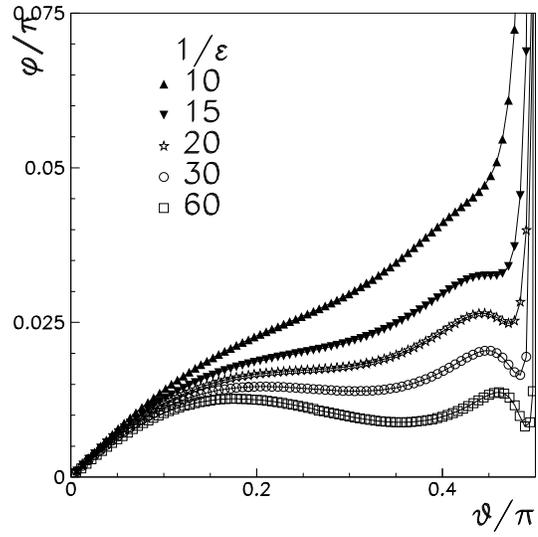} }
  \caption{The relationship between $\phi$ and $\theta$ for different 
   anisotropies, with $\kappa=50$.}
\end{figure}

\newpage
\begin{figure}
  \narrowtext
  \centerline{\epsfxsize=6.5cm
  \epsfbox{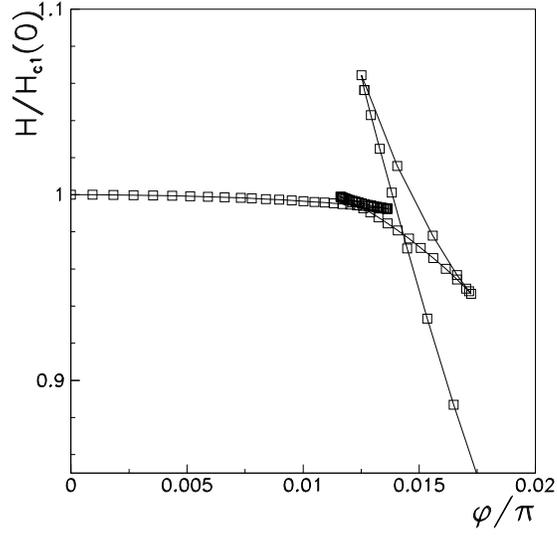} }
  \caption{The relationship between ${\cal H}$ and $\phi$ showing the 
   existence of two kinks in $H_{c_1}(\phi)$, for $\kappa=50$ and 
   $\epsilon=1/40$. This allows the orientation of the chains state 
   between two forbidden regions.}
\end{figure}

\begin{figure}
  \narrowtext
  \centerline{\epsfxsize=6.5cm
  \epsfbox{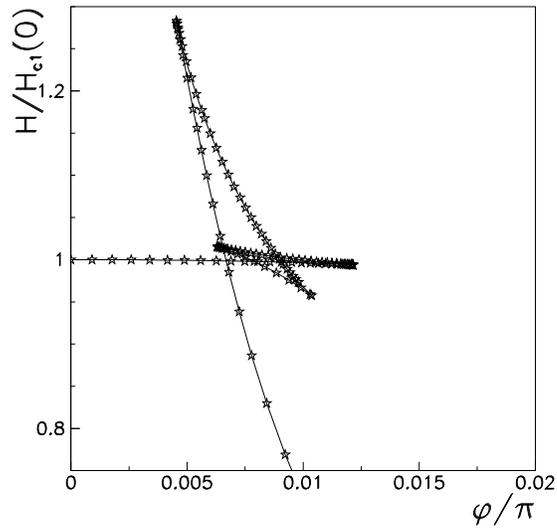} }
  \caption{${\cal H}(\phi)$ for $\kappa=50$ and $1/\epsilon=100$, 
   showing that at large anisotropy the second kink in $H_{c_1}$ 
   disappears.}
\end{figure}

\end{document}